# Deciphering Urban Morphogenesis:
# A Morphospace Approach

Vinicius Netto[1], Caio Cacholas[2], Dries Daems[3], Fabiano Ribeiro[4],
Howard Davis[5] and Daniel Lenz[6]

[1] Research Centre for Territory, Transports and Environment CITTA | Faculty of Engineering
FEUP | University of Porto, 4200-465 Portugal | vmnetto@fe.up.pt

[2] Graduate Programme in Architecture and Urbanism, Fluminense Federal University |
PPGAU UFF, Brazil

[3] Department of Archaeology, Vrije Universiteit Amsterdam, Netherlands

[4] Department of Physics, Federal University of Lavras | UFLA, Brazil

[5] Department of Architecture, University of Oregon in Eugene, US

[6] Institute of Architecture, Urbanism and Design, Federal University of Ceará | UFC, Brazil

**Abstract.** Cities emerged independently across different world regions and
historical periods, raising fundamental questions: How did the first urban
settlements develop? What social and spatial conditions enabled their emer-
gence? Are these processes universal or context-dependent? Moreover, what
distinguishes cities from other human settlements? This paper investigates the
drivers behind the creation of cities through a hybrid approach that integrates
urban theory, the biological concept of *morphospace* (the space of all possible
configurations), and archaeological evidence. It explores the transition from
sedentary hunter-gatherer communities to urban societies, highlighting fun-
damental forces converging to produce increasingly complex divisions of
labour as a central driver of urbanization. Morphogenesis is conceptualized as
a trajectory through morphospace, governed by structure-seeking selection
processes that balance density, permeability, and information as critical di-
mensions. The study highlights the non-ergodic nature of urban morphogene-
sis, where configurations are progressively selected based on their fitness to
support the diversifying interactions between mutually dependent agents. The
morphospace framework effectively distinguishes between theoretical spatial
configurations, non-urban and proto-urban settlements, and contemporary
cities. This analysis supports the proposition that cities emerge and evolve as
solutions balancing density, permeability, and informational organization,
enabling them to support increasingly complex societal functions.

# 1. Introduction

Cities are everywhere — an extraordinary spatial solution in which around 3% of the global territory is home to 57% of the world's population. In developed countries, cities are home to more than 80% of their populations (UNCTAD, 2023). However, how did we get to cities? What social and material forces led distinct societies in different times and regions to independently produce cities as a spatial solution for accommodating collective social life? Are cities universal or context-dependent? These questions address the creation of cities as a response to growingly complex societies and perhaps a part of their emergence. While non-urban cultures in hunter-gatherer societies continue to exist today, most others have taken the path of urbanisation, producing complex functional spatialities we call 'cities'. We can tentatively define a city as a densely populated and spatially structured settlement characterized by a high degree of social complexity, economic specialisation, and functional diversity. It supports a concentrated population through interconnected systems of infrastructure, governance, and services, enabling collective activities, production, and exchange. Key drivers usually identified in the literature include defence, social hierarchies, population scale, food production, commercialisation, and work specialisation (Cowgill 2004; Creekmore and Fisher 2014). Our approach will focus on specific social and material forces that drove urbanisation, triggered by an increasing complexity of the division of labour.

Once we account for cities as sociospatial solutions for changing patterns of collective life, new questions arise: What distinguishes cities from non-urban settlements in terms of form and function? Why did both common structures and diversity of forms arise? These questions are classic subjects of urban morphogenetic studies (e.g. Alexander, 1964; Hillier and Hanson, 1984), a branch of urban morphology. 'Morphogenesis' originated in biology as the process through which cells, tissues and organisms acquire dimension and form (Gilmour et al., 2017). In urban studies, morphogenesis addresses the creation of the specific spatialities of cities (cf. Conzen, 1988). One way to address this question is by deploying another concept originally developed in biology: 'morphospace' (Raup, 1966). A morphospace is the space of all possible biological configurations. This concept is interesting in the context of cities for several reasons. Among them, it allows us to think counterfactually about the possibilities of form for human settlements, what morphological paths were explored by human spatial cultures, what possibilities exist but have not been explored so far – and the reasons for not having done so.

The concept of morphospace will allow us to look into the passage from all possible forms in a bidimensional abstract space, to all possible forms explored by *human* settlements to all possible *urban* forms. The morphospace approach requires us to analyse features of morphology and morphogenesis from both an evolutionary and structural perspective, including how urban morphology differentiates itself from other types of human settlements. Spatial economics addresses this question, seeing agglomeration as the spatial solution (e.g. Fujita et al., 1999). Still, agglomeration is a general spatial property that may be achieved through a virtually endless number of spatial arrangements. A morphospace approach can show that most of the potentially attainable forms of spatial organisation simply could not work as cities — because cities are likely to display specific spatial properties. Furthermore, these properties may delineate the position of cities in the morphospace. Cities need not only to generate density in agglomeration but also allow people to move and navigate spaces around them if they are to interact. A morphospace has to materialize distinctive and related spatial properties that can characterise and distinguish the vast range of spatial configurations, ideally spanning from all possible forms to non-urban and urban settlements.

We shall focus on three fundamental properties that cities might have emerged to balance: density, permeability, and information in urban form. We will argue that these three properties are sufficient for describing the differences between any configuration, including non-urban and urban settlements. We develop measures of such properties and apply them to a sample of configurations, from theoretical spatial distributions to actual human settlements, proto-cities and urban configurations. These properties delineate a morphospace for human settlements, identifying clusters of morphologies that emerge from different parameter sets. In the spirit of the concept in biology, the morphospace will allow us to see the space that cities tend to occupy, the positions that different types of non-urban settlements tend to occupy, and the spaces that do not seem possible for human settlement.

In short, the paper introduces a framework for deciphering urban morphogenesis through the concept of morphospace. First, it briefly considers the creation of cities as a response to the growing functional complexity of certain societies. Second, it develops a means to analyse spatial conditions for the emergence of urban forms. By synthesising insights from urban theory, biology, and archaeology, we seek to elucidate how cities evolved through trajectories from the space of all possible spatial configurations to those actually found in human settlements and, subsequently, urban settlements. Finally, the paper discusses the implications of the morphospace approach for understanding urban morphogenesis. We shall begin with a brief review of approaches to urban morphogenesis.

## 2.    Literature review: approaches to urban morphogenesis

Cities emerge as adaptive spatial solutions facilitating social interaction, trade, cultural development, and governance. They often serve as centres for innovation, resource allocation, and decision-making. This definition reflects key attributes commonly found in cities:

- Density and Scale: Higher population density and spatial concentration relative to surrounding areas.
- Social Complexity: Diverse social structures and networks often lead to hierarchical social organization (Daems, 2021).
- Economic Specialization: Division of labour and specialization of economic functions that go beyond subsistence, fostering markets and complex economies.
- Functional Diversity: A range of functions, including residential, commercial, administrative, and recreational, that support collective life.
- Infrastructure and Governance: Organized systems of roads, water supply, sanitation, public services, and governance structures, which manage resources and population needs.

Cities evolved by filling the available space for them in different ways, in different densities and using different patterns to bring energy in terms of people and materials that allow their constituent parts to function (Batty, 2013). A pioneering approach to such dynamics is diffusion-limited aggregation models (DLA; e.g. Batty and Longley, 1989; figure 1a). DLA models describe urban growth as a process similar to particle aggregation in physical systems. Initially developed in physics to explain how particles adhere to form clusters, DLA models simulate urban expansion by envisioning cities as clusters that grow as "particles" (representing people, buildings, or resources) diffuse through space and attach to existing urban areas. In this model, growth occurs when new particles or elements join the periphery of the existing cluster, with attachment probability influenced by proximity. This leads to organic, branching, or fractal-like patterns that mimic the uneven, incremental, and self-or-ganizing growth seen in some cities. The resulting form often includes densely connected cores with branching structures extending outward, resembling urban sprawl or informal settlements. DLA models are valuable for understanding how cities evolve without centralized planning, particularly when natural diffusion processes, local interactions, and limited resources constrain growth. These models reveal how urban forms can emerge from local decision-making and random movements, demonstrating how spatial structure and connectivity emerge naturally from decentralized growth dynamics. However, recalling Alexander's (1966) seminal work, a city is not a tree. Cities do not have dendritic branches of poorly connected built forms or streets. They are spatial networks of high circuitry, creating permeability that allows mobility in different directions at different positions in the structure.

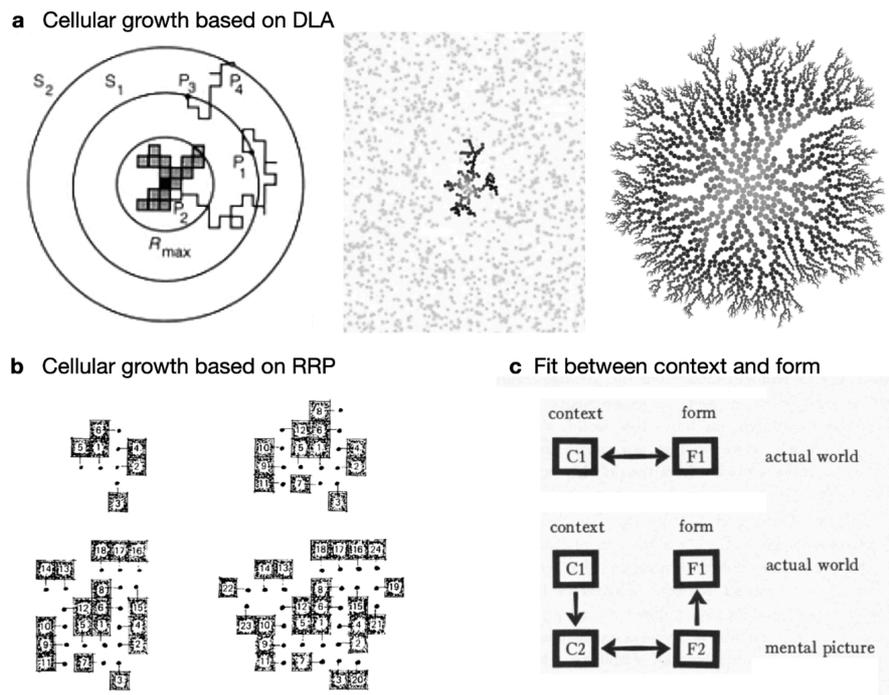

**Fig. 1.** Approaches to urban morphogenesis: (a) Dendritic cellular growth based on diffusion-limited aggregation (DLA, Batty and Longley, 1989; Bourke, 2023). (b) Cellular growth based on built-form adjacency, continuity of open cells and restrictions on random processes (RRP) (Hillier and Hanson, 1984). (c) Fit between context and form (Alexander, 1964).

Urban structure can also be conceptualised regarding restrictions on an otherwise random process, a model including randomness and order. Settlements may be generated through local rules, as degrees of restriction to random processes leading to well-defined global forms (Hillier and Hanson, 1984:11). This is a powerful principle capable of generating adjacencies between cells of open space or cells of built form as early stages of emerging spatial microstructures (figure 1b). However, how local restrictions lead to micro-structures, namely the critical emergence of the urban block as 'rings' of buildings, along with the macrostructures in the form of block systems arranged symbiotically with street networks, remains unclear in such representations. Also, we have to test how far the restrictions on randomness go (as ways to reduce the entropy of spatial arrangements) in leading to urban structures.

Christopher Alexander (1964, 2002) introduced key morphogenetic concepts to urban design. "Goodness of fit" refers to how well a spatial configuration addresses the functional, social, and environmental needs of its users, minimizing conflicts or "misfits." This idea is expanded through "fit-misfit," where cities evolve by resolv-ing spatial mismatches to achieve harmony. The "trial and error" process is central to urban evolution, where iterative changes improve configurations by adapting to real-world challenges. "Structure-preserving transformations" emphasize maintaining the core integrity of forms during changes, allowing cities to adapt and grow while retaining their foundational elements and functional identity, seeking to harmonize spatial configurations with social and material needs. However, Alexander did little to render such concepts applicable to systemically analysing urban form or its evolution.

More recent approaches include a density-based model of urban morphogenesis (Raimbault, 2018) of a variety of generated urban shapes — from a very diffuse urban configuration to a semi-stationary polycentric urban configuration, intermediate settlements (peri-urban or densely populated rural area), and rural areas. Mapping processes of accretion to urban form — based on street networks or/and buildings (e.g. Dovey et al., 2020; Kamaliour and Iranmanesh, 2021). These works usually focus on the morphological evolution of cities in specific historical periods as accretion processes to urban form rather than the original social and material forces driving the city-creation process. In turn, typological analyses (e.g. Panerai et al., 2004) mainly focus on the morphological evolution of cities in specific historical periods. They do not pose fundamental questions about the original social and material forces that led to the creation of cities independently in different regions and historical moments.

## 3. Context: Why were cities formed?

The emergence of cities across diverse societies and cultures raises fundamental questions about why and how urban settlements took shape independently in different regions and historical contexts. Are cities a spatial solution to accommodate shifting patterns of collective life? Do they represent a convergence of deeper social and material forces? Understanding the drivers behind city formation requires examining the conditions that enabled the first urban settlements, from population increase to innovations in food production, commerce, defence, and governance. This section explores the social and material forces that influenced early urbanisation, analysing whether cities arose primarily as a functional response to complex social needs or as a byproduct of unique, context-dependent factors. By tracing the origins of cities across different regions, we aim to uncover the universal and context-specific dynamics that made cities a preferred structure for human habitation and organisation. Several potential reasons for the creation of cities have been explored.

- *Defence and Protection:* Cities provided a centralised, fortified space where people could seek safety from external threats, such as invasions or conflicts, allowing populations to defend themselves collectively (Lambert, 2002; LeBlanc, 1999).
- *Power and Social Hierarchies:* Cities enabled the concentration of power and facilitated the establishment of social hierarchies, with centralised governance structures that could manage resources, enforce laws, and maintain order (Ur, 2014; Wright and Johnson, 1975; Goulder, 2010).
- *Increase in Population Scale:* Growing populations required new forms of organisation and infrastructure to support dense living arrangements, leading to the emergence of cities that could accommodate and manage large numbers of people (Boyd and Richerson, 2005; Smith, 2019; 2023; Smith and Lobo, 2019).
- *Need to Increase Food Production:* Cities often formed near fertile land or trade routes, enabling organised agricultural practices and food storage systems to support larger populations (Jacobs, 1969; Adams, 1981; Goulder, 2010).

- *Commercialization:* Cities became hubs for trade and economic exchange, where markets and merchants could operate efficiently, facilitating the flow of goods and ideas across regions (Trigger, 1972; Bairoch, 1991).
- *Work Specialisation:* Cities fostered a complex division of labour, enabling individuals to focus on specialised tasks, which increased productivity, interdependence, and innovation, fuelling urbanisation (Fisher and Creekmore, 2014; Southall, 1973; Trigger, 1972; Patterson, 2005).

Among others, these factors may have combined to create favourable conditions for cities, reinforcing the urbanisation process in unique but interconnected ways. We shall look more closely into the relationship between work specialisation and urbanisation through the lens of increasing complexity in cooperation and the division of labour (Cooper and West, 2018; Cooper et al., 2021) and increasing functional diversity in an urbanising economy (Trigger, 1972; Hanson et al., 2017). Finding evidence of the relationship between population and functional diversity is a challenging empirical issue. However, the relationship between division of labour, specialisation and functional diversity was found in ancient Roman cities (e.g. Kaše et al., 2022; Hanson et al., 2017). We hypothesise that there are several implications of the division of labour for the historical emergence of cities:

- **Increasing interaction and mutual dependence between specialists** (social vector): As societies become more complex, individuals specialise in specific tasks and professions, from crafts and trade to administration and governance. This specialisation creates a social network of interdependent roles, where individuals rely on the skills and outputs of others to fulfil their own needs. This mutual dependence intensifies interactions between specialists, fostering a social environment that requires frequent collaboration, exchange, and communication. With their connected structure, cities naturally support this increasing interaction and cooperative, interdependent relationships (Jacobs, 1969; Fujita et al., 1999; Bettencourt et al., 2014; Ortman et al., 2014; Smith, 2019).
- **Need to reduce the distance between specialists** (material condition): The division of labour creates the need for proximity, as specialists benefit from being close to one another to facilitate efficient exchanges and collaborations. Reducing the physical distance between professionals lowers transaction costs, improves the flow of goods and information, and supports timely interactions essential for a functioning specialised economy. Cities fulfil this material condition by concentrating people and resources within a limited geographic area, minimising travel time and encouraging an interconnected network of specialists who can efficiently interact and trade within the same spatial context (Storper and Venables, 2004).
- **Increasing density** (material response to increasing interaction and mutual dependence): As the division of labour and population scale grow, the density of settlements naturally increases. This density is a practical response to the need for close interactions in a limited space and accommodating the rising number of people required to support a dense network of co-dependent agents in an increasingly complex economy (Fletcher, 2007; Glaeser, 2011; Ribeiro et al., 2024).
- **Need for mobility for mutually dependent specialists** (functional response to the challenges above): The spatial separation between where people live and work is a byproduct of increasing specialisation and cooperation in work. Specialists, therefore, need efficient mobility to navigate between spaces, allowing them to connect with others and access resources. This mobility requirement shapes the structure of cities, prompting the development of permeable block systems, pathways and street networks that facilitate the movement of people, goods, and information within a complex spatial system that accommodates the division of labour and collective interdependence.

We can summarise such a set of social forces and material requirements in societies with a growing division of labour and functional diversity:

- Increasing interactions.
- Density as a means to generalise proximity.
- Mobility, which requires spatial permeability within settlements.

Societies with growing divisions of labour require spatial configurations that satisfy the need for proximity between specialists in increasing numbers (density) and the mobility of those specialists around the barriers that buildings, once built, place in a landscape. When both increase, there is a tendency to contradiction and conflict (figure 2). How can we spatially resolve these coexisting forces?

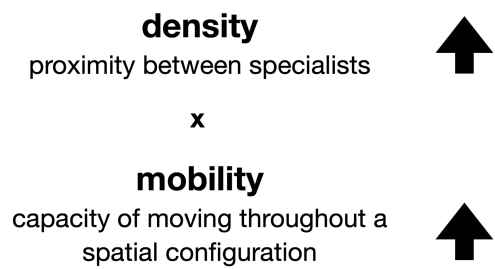

**density**
proximity between specialists

x

**mobility**
capacity of moving throughout a
spatial configuration

**Fig. 2.** Social forces and material conditions: in scenarios where both density and mobility are expected to increase, as in situations of growing populations of mutually-dependent specialists, both conditions can easily be conflicting.

## 4. From all possible forms to urban form: Morphogenesis as a trajectory across morphospace

Different spatial cultures managed to solve the trend of conflict between the increasing need for density and the increasing need for mobility through incredibly diverse morphological ways. Looking into such ways can help us identify spatial properties, entities, and innovations that may express and support such social and material requirements, characterise cities, and be found in different urban forms — and what the differences are between them and other types of human settlements. So, firstly, we must acknowledge the diversity of settlement forms (Figure 3). How did such a diversity of settlement forms arise? Can such different spatial forms support complex modes of collective life? The complexity and technology of hunter-gatherer societies have been extensively studied (Hoffecker and Hoffecker, 2018 – see settlements (a) and (b) in Figure 3).

These societies display fundamental characteristics that structure social organization, resource allocation, and survival strategies (Solich and Bradtmöller, 2017). The division of labour is less defined, as people had more diverse roles depending on the context than in agricultural or industrial societies. Tasks such as hunting, tracking, crafting tools, shelter-building, plant gathering, processing food, and caring for younger children are generally divided by gender and age and frequently require group coordination. People may switch roles based on necessity, environmental conditions, or personal skill sets, with individuals often possessing a wide array of skills rather than highly specialized ones. The need for adaptability discourages strict specialization and is essential for survival in environments where resources can be unpredictable and where rigid specialization could be detrimental to group resilience. Population density, which is closely linked to increasing specialization, mutual dependence, and interaction, is difficult to achieve in spatial arrangements character-ized by low building density and large distances between housing or production units, as typically seen in such societies.

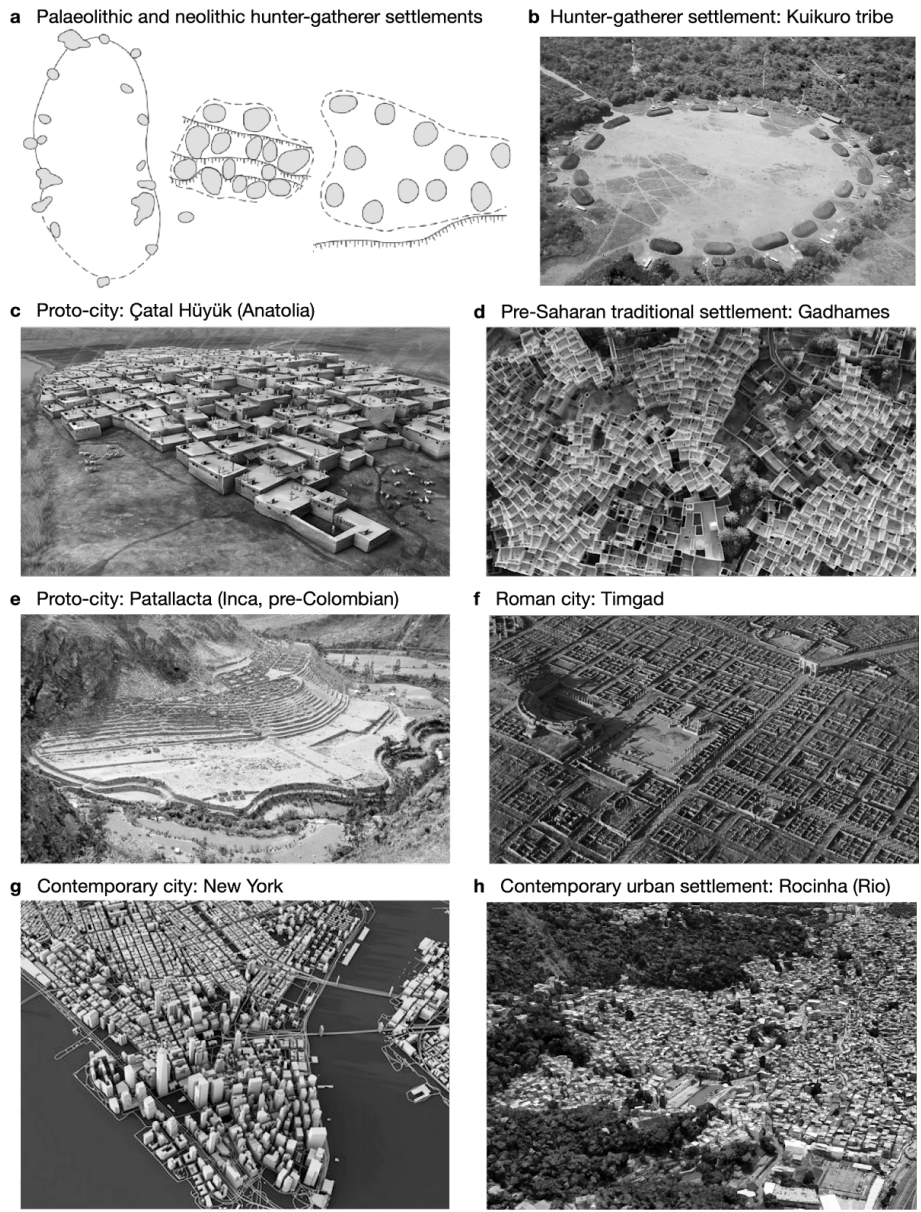

**a** Palaeolithic and neolithic hunter-gatherer settlements
**b** Hunter-gatherer settlement: Kuikuro tribe
**c** Proto-city: Çatal Hüyük (Anatolia)
**d** Pre-Saharan traditional settlement: Gadhames
**e** Proto-city: Patallacta (Inca, pre-Columbian)
**f** Roman city: Timgad
**g** Contemporary city: New York
**h** Contemporary urban settlement: Rocinha (Rio)

**Fig. 3.** Illustrating the spatial diversity of human settlements: (a) Kostienki I; Prepottery Neolithic huts from Level II of Nahal Oren, Israel (ca. 8000-6000 BCE) and Çayönü, Near Eastern pre-pottery Neolithic (8000-6000 BCE). (b) Kuikuro tribe, Alto Xingu (contemporary Amazon, Brazil). (c) Çatal Hüyük, Southern Anatolia, present-day Türkiye (7500-6400 BCE). (d) Gadhames, pre-Roman oasis, present-day Libya (4000 BCE). (e) Patallacta, pre-Columbian Inca society, present-day Peru (800 AD). (f) Timgad or Thamugas, Roman city, present-day Algeria (100 AD). (g) Manhattan, New York City (United States). (h) Rocinha, informal urban settlement in Rio de Janeiro (Brazil).

However, density is a spatial property that can be achieved through many configurations. What exactly spatially distinguishes non-urban from urban settlements? The concept of morphospace may help clarify those processes. Resembling the definition of *phase space* in physics, morphospace means 'the space of all possible configurations'. It was introduced by Raup (1966) to explain the differences between gastropod shells. In Figure 4, only regions A, B, C, and D are occupied by shells of actual species, of which examples are illustrated. Other regions describe shells that are theoretically possible but are not found in nature.

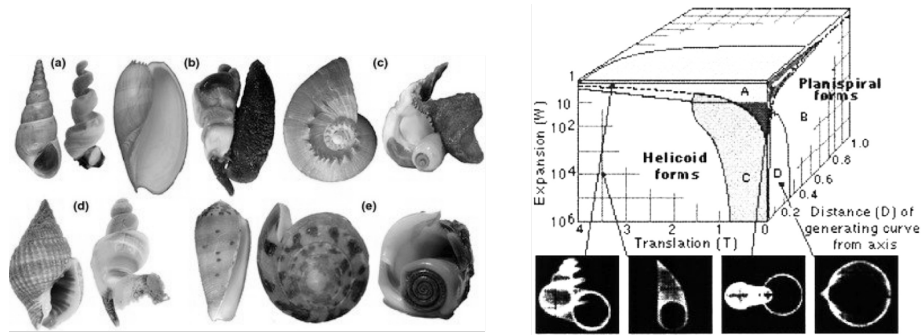

**Fig. 4.** Raup's morphospace for gastropod shells (Raup, 1966).

What would the morphospace of human settlements look like? It should map all possible configurations and allow us to spot the differences between different settlements, from hunter-gatherers' settlements to cities. Nevertheless, considering morphospace as a space of infinite configurations is challenging. A potentially helpful property of 'morphospace' is its natural affinity with a concept familiar in rigorous studies of dynamic systems in physics: 'ergodicity'. An ergodic behaviour means that a system visits all possible combinations of available elements. The ergodic hypothesis is that over a sufficiently long time, a system will explore all accessible configurations with equal probability. The gas particles within a box provide the classic example: in a sufficiently long period of time, particles will occupy every possible position and velocity combination consistent with the constraints of the system (such as energy and volume). As the gas evolves, entropy tends to increase as particles can move free from specific temporal and spatial trajectories, and their distribution is as random as possible. This is a critical concept in seeing evolutionary phenomena, even though purely ergodic behaviour is impossible above the level of molecules (Kaufman, 2000).

This is undoubtedly the case in human exploration of different spatial configurations. Spatial morphogenesis does not allow for exploring most possible configurations. Spatial systems have materiality: they are produced with significant human effort and material resources. They show rigidity and permanence – therefore, resistance to change. They have a history and path-dependent trajectories: previous states determine in part later ones, which can lock in specific patterns and be historically shaped into different spatial cultures (Netto, 2017). Humans create spatial configurations by visiting and experimenting with various configurations, including in imagination, but do not examine and exhaust every possibility, let alone equiprobable possibilities. At best, humans can explore possibilities in a 'quasi-ergodic' way while imagining, seeking and experimenting with configurations that could fit that population or society.

Spatial morphogenesis is not an ergodic process. Yet, we still have to consider that it can reach any possible configuration in principle — so which ones can become cities? We can imagine differences between a "morphospace of all possible configurations" (morphospace 1), a "morphospace of all possible human settlements" as configurations suitable for collective life (morphospace 2), and a "morphospace of all possible urban settlements" as configurations suitable for urbanising societies with a growing division of labour. They all seem to contain limitless possi-

bilities – yet morphospace 1 is larger than morphospace 2, which is larger than morphospace 3:

**morphospace 1** > **morphospace 2** > **morphospace 3**
(all possible configurations)    (all possible *settlements*)    (all possible *urban* settlements)

We propose understanding the evolutionary stages of morphogenesis as a trajectory across morphospaces. We can do so by exploring what Alexander (2003) called "walk in configuration space": how does the reduction from morphospace 1 to morphospace 2 to morphospace 3 come into being in morphogenesis (Figure 5)?

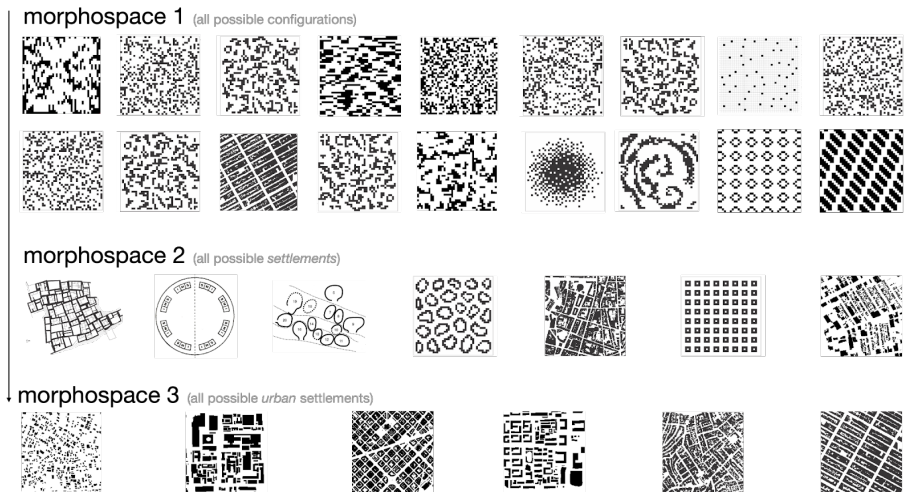

**Fig. 5.** Walking in configuration space: a schematic illustration depicting the transition from the space of all possible bidimensional configurations (morphospace 1) to the subset of configurations that can support human settlements (morphospace 2) and further to the subset possibly suited specifically to urban settlements (morphospace 3).

What guides this trajectory? Does it involve the selection of configurations made by those building the original settlements? If so, how does this selection occur? Is it shaped by trial and error or by adapting configurations over time? What rules, whether conscious or unconscious, guide the choice of forms during spatial production? Additionally, what spatial characteristics must arrangements have to become cities eventually? Processes of trial and error and the fit between context and form (Alexander, 1964) – or, more precisely, the fit between interaction and form, or even the adaptation of form to interaction – can guide the selection and shaping of configurations that meet the need for proximity among specialists as their numbers (density) increase, as well as their need for mobility despite the barriers that buildings, once constructed, create in the landscape. Much like the potentially conflicting relationship between increasing spatial density and agents' mobility, the alignment between interaction and form is fundamentally a spatial problem. In fact, the issue of density and mobility is a specific case of this broader challenge. So, how do we resolve it?

Agglomeration and density are essential for supporting scale and functional diversification within a growing division of labour (Jacobs, 1969; Fujita et al., 1999), yet they alone are insufficient. Achieving a balance between social dynamics and material conditions requires not just density, but density combined with permeability. Permeability is a spatial condition that enables interaction—a quality that ensures a settlement provides mobility for its inhabitants. While permeability can be achieved through open spaces surrounding buildings, this material attribute alone is not enough. Actual permeability must extend beyond local spaces, facilitating movement across all parts of a settlement. How can this be achieved? We propose that density and permeability can be simultaneously achieved through structures capable of supporting densification while allowing full mobility and connectivity across different areas of a settlement.

## 5. Method: Morphogenesis as a path to structure

Across the range of urban cases throughout history — from Neolithic proto-cities like Sha'ar HaGolan (ca. 6400–6000 BCE) in the Jordan Valley (Garfinkel, 2006) to Patallacta, a pre-Colombian Inca settlement (800 AD), Timgad, an ancient Roman city in present-day Algeria (100 AD), and contemporary cities (Figure 3) — we observe common features amid the diversity of forms. Despite variations, these urban forms consistently exhibit aggregations of buildings organized into blocks of various shapes and sizes, connected by networks of open spaces or streets. What accounts for both this shared structure and the diversity of forms? A morphospace approach suggests that the diversity explored by different spatial cultures arises naturally within a seemingly limitless range of possibilities – constrained by specific material, social and contextual conditions. Thus, the first significant challenge lies in understanding what those different spatial cultures share, each independently developed across various world regions and historical periods. We argue that one spatial feature shared among these urbanising cultures is that they shape agglomeration through basic morphological elements – namely, urban block and street systems.

We have approached morphogenesis as a trajectory across morphospaces. Now, we can explore a more well-known facet of morphogenesis. A concept originating in biology, morphogenesis means the process by which systems' form, structure, and organization arise during development. It involves the generation, differentiation, and spatial arrangement of composing elements – like cells and tissues. Morphogenesis is a dynamic process that requires the coordination of cellular growth into pathways and systems. The process closely resembles Alexander's (2003) "structure-preserving transformations" and another concept, "harmony-seeking computation", which describes how forms evolve through an iterative, rule-based process that seeks an optimal configuration, or "harmony", within a set of constraints – where forms self-organize and adapt to achieve spatial and functional balance. For cities, this computational metaphor underscores how complex urban forms can emerge naturally from simple rules, gradually creating patterns that resonate with human needs and environmental context.

The teleological search for optimisation in Alexander's concept can be avoided, but the need for structuration remains – from biological and social systems (Maturana and Varela, 1972; Luhmann, 1995) to cities. As complex systems, cities historically exhibit self-organizing properties, developing structures and patterns spontaneously from the interactions between components within the system (see, e.g. Batty, 2013; Dries, 2021). Seeking to understand how cities achieve this remarkable feat, we approach urban morphogenesis as a self-organizing process through the idea of a "structure-seeking growth" guided by social and material requirements. In the case of social phenomena, systems like the economy can achieve coherent structures and functional organization through feedback loops (Arthur, 1994), positive and negative regulatory mechanisms, and other interactions. The idea of morphogenesis as adaptive, structure-seeking growth emphasizes the interactions and feedback loops that drive the emergence of functionally differentiated, specialised structures in complex systems.

One way of understanding how structures can be achieved in growing settlements is through the concepts of entropy and information. First, as distributions of entities, spatial configurations have different levels of randomness and order. These levels can be assessed by measures such as entropy and information. There are two main branches of definitions of entropy, namely in physics (e.g. Boltzmann and Gibbs entropies) and information theory (Shannon entropy). Essentially, entropy can be seen as a measure of probability. When the probability of an event, position, or frequency of occurrence is evenly distributed over an area or throughout a space or unit of time, the level of uncertainty in the system and randomness in its distribution tends to be very high. It is very hard to predict the behaviour of such a system. By measuring entropy, we may assess the probability of an arrangement as a whole or the probability distribution of events (e.g. cells) occurring within a spatial arrangement (Netto et al., forthcoming) (Figure 6).

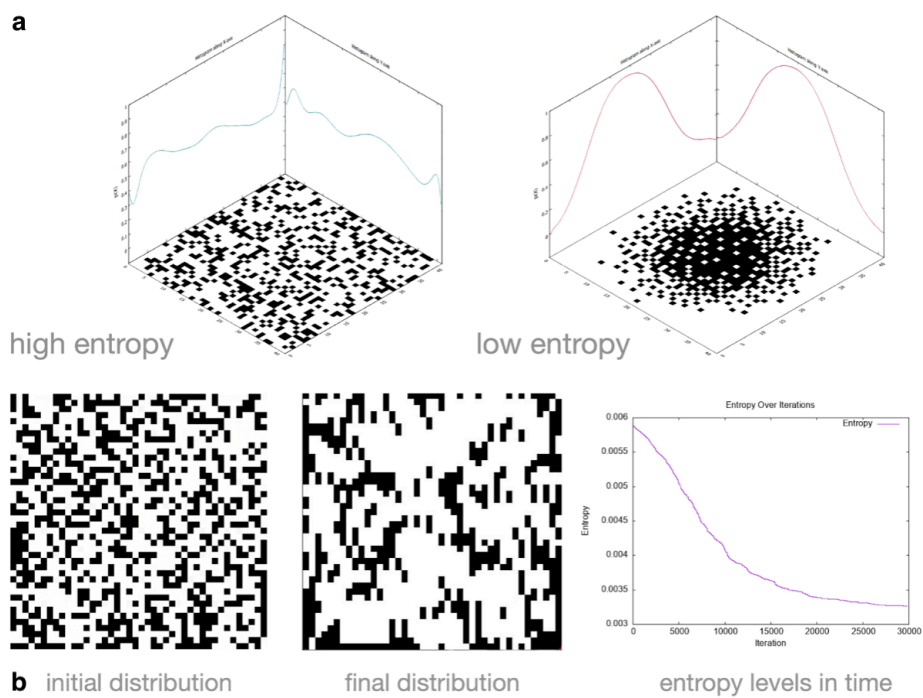

**a**

high entropy          low entropy

**b** initial distribution          final distribution          entropy levels in time

**Fig. 6.** (a) Two different cell distributions: a random, high-entropy distribution where cell positions find no pattern as shown by the blue curves, implying that it would be hard to pre-dict where the next cell would occur; and a low entropy, clustered distribution of cells with a dropping density according to the distance from the centre. The red curves predict the new cells' position in such a distribution. (b) Simulation of a distribution with falling levels of entropy over time. Despite increasing order, the final distribution neither reaches a visible structure nor resembles a human settlement (Netto et al., forthcoming).

We can see that distributions with internal correlations and consistencies increase the frequency of certain positions, like a higher number of aligned cells in the final distribution in Fig. 6b. However, our computational experiment in entropy reduction shows that we need more than restrictions on randomness or pure entropy reduction to get to cities. Even a low entropy distribution can be far from distributions found in urban settlements. Furthermore, urbanising societies probably cannot visit a large number of possible states. Different states do not have an equal probability of being visited by urbanising cultures. Most configurations have too high entropy to fit heightened levels of social interaction and the need for mobility thus engendered. When creating cities in different regions, societies needed to look for more 'rare' configurations: those able to fit the material requirements of interaction in complex divisions of labour. Even when considering rare distributions containing internal correlations, only a comparatively small set of cases could generate density through proximity and contiguity between cells while keeping permeability and could evolve into something like a city. The very *formation of ring-shaped micro-structures in the form of urban blocks* is a highly unlikely event. It requires selection. In morphogenesis, changes in configurations will be tested or inferred and discarded as unfit. For instance, highly dense distributions with high levels of disorder in the distribution of cells (high entropy distributions) might get in the way of people's ability to move in any direction.

In Shannon's (1948) problem of communicating strings of symbols through circuits, surprises may be understood as information. In spatial navigation, humans search for correlations, alignments, and visual cues to understand geographical environments, recognising regularity in the frequencies of spatial events in the built environment as they anchor cognitive systems (McNamara et al., 2003; Ekstrom et al., 2019). Hence, our approach resembles the view of information in organized structures and patterns seen in physics (Davies, 2015; Hidalgo, 2015). However, reducing entropy is not enough. We can have low entropy and poor permeability and density. Permeability is not enough: we can have permeability and poor large-scale structure. Societies with growing divisions of labour and heightened functional mutual dependence require some form of structure in densities to allow movement. They need density, permeability, and reduced entropy by generating local-to-global structures. In short, they need to *increase information along with density and permeability.*

Information functions as both an intrinsic element of systems and a method to analyse and understand structures within them. Increasing information indicates the presence of consistencies and regularities, which may lead to structures. The structures that interest moving agents allow for both density and mobility through permeability between buildings. Information combined with permeability, found in the continuity between cells of open spaces, bridge the microstructure of the urban block into large-scale correlations at a distance found in the macro-structure of urban block systems and street networks. In other words, they allow large-scale structures to emerge, bringing deep mobility within the system. *Information and permeability are ways into structure.*

### 5.1. Measurement

We selected specific measures to estimate density, permeability and information in cellular distributions, as dimensions of morphospace.

a. **Density:** a standard measure of form in urban morphology based on the ratio of built form area to the total area of a settlement or distribution.
b. **Permeability**, in turn, is a more complex issue. Existing approaches have limits regarding measuring features, whether based on urban blocks as spatial entities (e.g. perimeter, area, etc.) or streets (length, number of intersections, closeness centrality, etc.). We measure permeability as the product of the perimeter and area of each block divided by the total area of open spaces in the area under analysis (as opposed to the total area of urban blocks in Pafka and Dovey, 2016), normalized to give increasing values as permeability increases. This measure is sufficient to grasp permeability at local scales, but it is limited to grasp long-range permeability in global structures. We compensate this limitation by incorporating a measure geared to grasp consistencies across scales — information.
c. **Information:** our method estimates the probabilities of distinct arrangements of cells within blocks of 16 cells by counting their frequencies in bidimensional cellular configurations based on Shannon's entropy formula (see Brigatti et al., 2022; Netto et al., 2023 – Figure 7b), i.e. how many times every possible configuration of cells within moving blocks appears in areas under analysis. We minimised the effect of low and high densities over the probability of open or built cells by counting the frequencies of arrangements in the interfaces between built form and open spaces, ignoring homogeneous blocks of 16 cells.

We estimated *density* in a cellular grid representation of a bidimensional distribution of buildings and open spaces (building footprints) through:

$$\text{De} = \frac{\text{BFc}}{C_T} \tag{1}$$

where *De* corresponds to density, *BFc* is the total number of built form cells in the distribution and $C_T$ is the total number of cells. *Permeability* is measured through:

$$\text{Pe} = \sum_{i=1}^{n} P_i \times \frac{A_i}{A_O} \tag{2}$$

$$\text{nPe} = \frac{\text{Pe}}{\text{Pe}_{\max}} \tag{3}$$

$$\text{iPe} = 1 - \text{nPe} \tag{4}$$

where *Pe* quantifies the area-weighted average perimeter of urban blocks or built forms relative to the total area of open space ($A_o$) within the analysed settlement area. It is a measure of how the perimeters of blocks and built forms interact with open spaces. $P_i$ and $A_i$ are the perimeter and area of each block *i*, and *n* is the number of urban blocks or buildings (Equation 2). Equation 3 normalises *Pe* by dividing by its maximum value for bidimensional cellular configurations, $Pe_{max}$ — for simplicity, given by a distribution with a single built form cell over an open cellular field. *iPe* provides a normalised, intuitive metric for permeability that scales *Pe* to a relative range between 0 and 1 (equation 4). As iPe increases, it indicates higher permeability in the spatial configuration. Higher permeability means that (i) open spaces are less obstructed by built forms, and (ii) movement across the settlement is more facilitated. Finally, we estimate *information* based on Shannon entropy:

$$H = - \sum P_x \log_2(P_x) \tag{5}$$

$$\text{nH} = \frac{\text{H}}{\text{H}_{\max}} \tag{6}$$

$$\text{I} = 1 - \text{nH} \tag{7}$$

where *H* is the entropy of blocks of cells and the sum runs over all the possible *x* combinations of cells, $P_x$ is the probability of occurrence of combination *x*. Equation 6 normalises H by its maximum value for bidimensional cellular configurations, $H_{max}$, given by any distribution with full randomness and equal probability of finding built form cells or open space cells over the grid. Equation 7 provides an intuitive measure of spatial information considering navigation in settlement configurations, where *I* increases as normalised entropy *nH* decreases.

We applied these measures over representations of the spatial distributions of sampled settlements. For measuring *permeability*, we avoided the issue of isolated buildings surrounded by open spaces within urban blocks potentially read as permeable areas when they may not be by using the contours of blocks as collections of plots in their interface with public open spaces. We used maps available in Open Street maps with such representation (Figure 7a). For measuring *information*, we prepared our sample by extracting building footprints in sections of cities from the Google Maps service and archaeological data. Sufficiently large scales matter for grasping correlations, a necessary condition for correctly estimating entropy (see Netto et al., 2023). Cities and non-urban settlements can vary significantly in scale. We attempted to minimise this issue by applying the same grid structure regarding the number of pixels in representations. Images underwent a re-sizing process for producing sections with 9,000,000 cells. They were converted to a monochrome system and then into a matrix of $3000 \times 3000$ cells with binary numerical values (0 for open-space cells and 1 for built form cells). In the case of contemporary cities, we chose geographic areas of 9,000,000 m2, sufficient to represent the spatial characteristics of areas regarding the configuration of buildings, urban blocks, and open spaces following the same resolution and scale. The algorithm scans the map moving over neighbourhood cells, counting frequencies of cellular configurations within a window size n=16. Figure 7b illustrates amplified areas in red for two specific configurations. Rotation in built form does not affect results. In turn, *density* measurements for non-urban and proto-urban settlements smaller than 9,000,000 m² were based on the minimum perimeter encompassing all buildings (Figure 7d).

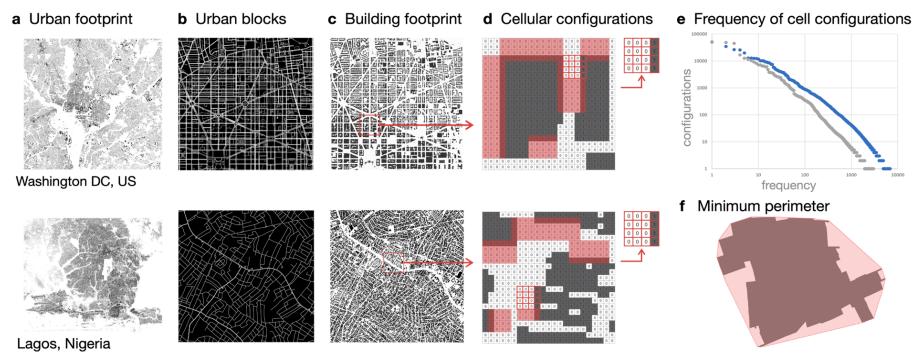

**Fig. 7.** Method: (a) Urban footprints of complete cities. (b) Permeability measurement in cities is based on urban blocks limits (Open Street Maps). (c) Information and density measurement is based on building footprints (Google Maps). (d) Information is computed as the algorithm scans built form maps counting frequencies of cellular configurations (n=16), with an example in red showing where it is found. (e) The total number of configurations n=16 found in Washington (grey line) is 2,989, and 6,825 in Lagos (blue line) out of 65,534 possible combinations. Washington consistently shows regular arrangements, which contrasts with the variety in Lagos. (f) Density measurement in non-urban and proto-urban settlements is based on the minimum perimeter encompassing all buildings.

## 6.    Empirical study: identifying settlements in morphospace

We shall explore the morphospace approach to investigate morphological variations and potential evolutionary transitions among human settlements. The aim is to examine how different morphologies relate to one another within the broader "space of possible configurations," identifying those regions within morphospace occupied by actual settlements and those that remain unoccupied. Each morphology is characterised by distinct parameter values, which may form clusters representing specific types of settlements. We analysed a dataset comprising 42 contemporary cities, 8 proto-urban settlements – including two sections of Çatal Hüyük (present-day Turkey) and six pre-Colombian Maya settlements (present-day Mexico) – and one example of a hunter-gatherer settlement, Bororo (present-day Brazil) and 5 theoretical configurations (Figure 8). The theoretical configurations (namely, perfectly ordered, random, tree-like and dispersed configurations) illustrate minimum and maximum values of density, permeability and information. These theoretical examples provide reference points within the morphospace, allowing us to position actual human settlements relative to them and better understand how they are distributed across this space.

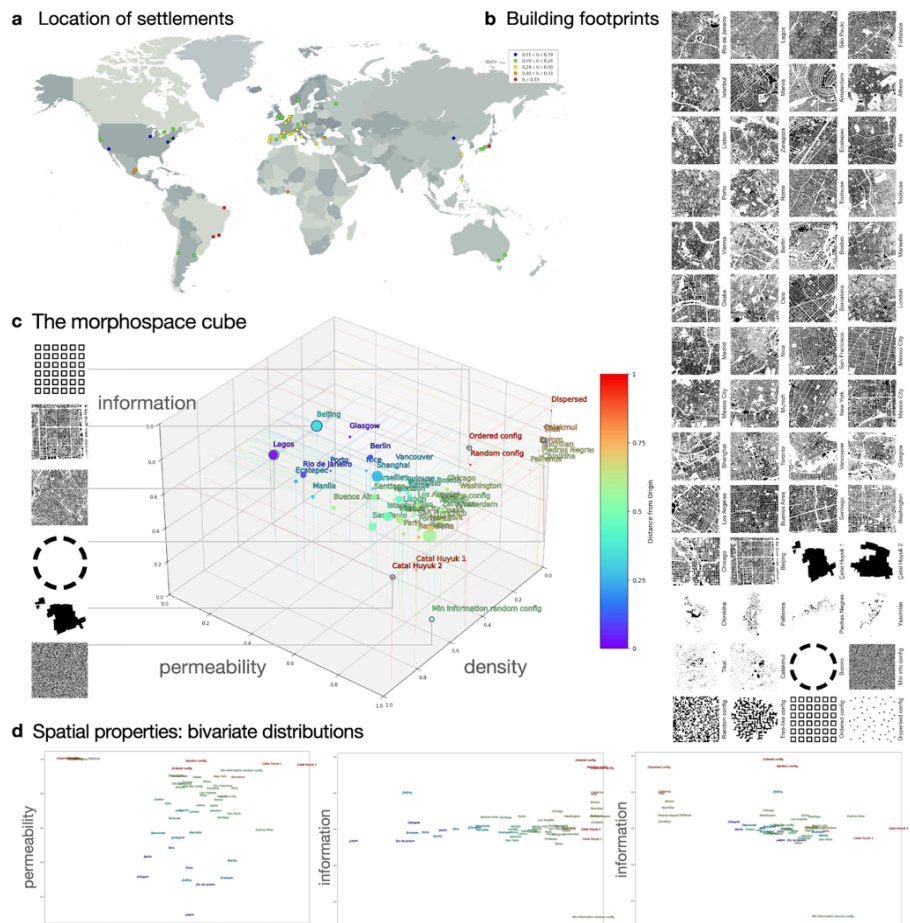

**Fig. 8.** Configurations in morphospace: (a) Location of real settlements in our sample. Colours show amounts of information found by our method in cellular configurations of built form, from blue (low) to red (high). (b) Building footprint distributions in urban and non-urban settlements. (c) The morphospace illustrates the positions of configurations according to information, permeability and density. Colours from purple to red indicate intensity. Dot sizes represent population sizes for all settlements derived from census data and anthropological and archaeological estimations. (d) Bivariate distributions of spatial properties.

The morphospace suggests certain clear differences between types of spatial configurations. First, contemporary cities are not homogeneously distributed across the cube but mostly cluster around a region of midrange of density, permeability and information values (green to purple dots in Figure 8c-d). Most urban sections analysed have densities between 0.35 and 0.6, permeabilities between 0.25 and 0.75 and information levels between 0.2 and 0.4. Some cities will show higher information levels, such as Chicago (US) and Beijing (China), due to more ordered layouts engendered by top-down traditional planning (see Brigatti et al., 2022), lower densities, such as Glasgow (Scotland), Berlin (Germany) and Vancouver (Canada), or low permeability, like Ecatepec (Mexico) and Manila (Philippines). Non-urban settlements, namely the Bororo village and Maya proto-cities form another cluster (yellow to orange dots) positioned in different regions of the cube, related to very low densities and high permeability. The two sections of Çatal Hüyük analysed show radically different levels permeability and density.

Contemporary cities occupy a well-defined region of the cube – especially in a narrow field within the range of possible information and density values for 2D cellular configurations (Figure 8d). Whole regions of the cube show positions that hardly could be occupied by human settlements, but find configurations such as the theoretical configurations illustrated here, displaying either very high or very low information, density and permeability.

## 7. Conclusions

This study addresses a fundamental question in urban studies: how did cities emerge independently across different regions and historical contexts, and what distinguishes urban forms from other types of human settlements? Our research sought to examine the underlying social and spatial drivers of urban morphogenesis, investigating the specific conditions and processes that guide the transition from settlements to complex urban structures undertaken in some regions. By conceptualizing urban morphogenesis as a "trajectory across morphospace," we adopted a hybrid approach that integrates concepts from urban morphology, biology, and archaeology. Our primary aim was to introduce a morphospace framework for analyzing urban evolution, which allows us to map the potential configurations of human settlements according to some unique spatial and structural properties characteristic of cities. This framework intended to explore:

1. *Balancing Density and Mobility:* We hypothesised that cities emerged by solving the increasing and often conflicting demands of density and mobility. These dynamics are essential to support complex divisions of labor and heightened social interactions, which are foundational for the functional diversification of urbanizing societies.

2. *Structure-Seeking Growth in Urban Morphogenesis:* The study highlights the role of "structure-seeking growth" in the non-ergodic processes of urban morphogenesis, where cities develop through progressive, path-dependent selection processes shaped by spatial constraints, social organization, and functional needs (cf. Netto, 2017). This approach challenges the notion of random or purely ergodic urban development.

3. *A Morphospace for Urban Analysis:* We propose a morphospace framework grounded in the interplay of density, permeability, and information. This analytical device distinguishes between theoretically possible spatial configurations, non-urban and proto-urban settlements, and contemporary cities. Interestingly, contemporary cities are concentrated in a specific region of the cube, particularly narrow regarding information and density values (Figure 8d).

4. *Testing Hypotheses in Morphospace:* The results of measures applied to sampled configurations in the proposed morphospace indicate that the hypothesis — that cities evolve as balanced outcomes of density, permeability, and informational organization — has withstood the empirical data without being falsified. This balance may represent a key feature of urban settlements, driving their ability to accommodate collective life and complex societal functions.

A key question arising from these findings is why cities consistently occupy a specific region within morphospace, avoiding positions above or below this range. While non-urban and proto-urban settlements may appear in different regions of morphospace, their positions remain limited compared to the full spectrum of possible configurations. The morphospace approach invites *new hypotheses*, such as the existence of a minimum threshold for cities to achieve sufficient density, permeability, and information. This threshold may include long-range correlations— or spatial structures—within their configurations, which are necessary to support the functional complexity of urban environments. A key aspect of such an evolutionary process is how the trajectory through morphospace may be driven by non-ergodic visits to possible configurations in the form of trials, errors (fit-misfit), and selection. For instance, once spatial innovations are introduced in urbanising societies and their settlements, structuring effects in the balance of density permeability are felt and selected for reproduction. They become part of the solutions that fit the interaction function, as cities self-organize through feedback mechanisms and adaptive growth. Localised communities would lock in on certain solutions — giving rise to path dependent reproduction and specific spatial cultures. These configurations are not necessarily the fittest but are *fit enough* to support mobility and interaction. We expect to investigate such processes further by (i) looking into the archaeological evidence on the formation of cities as a response to societies with growing functional complexity, and (ii) exploring morphogenetic computational experiments focusing on the spatial innovations that allow a range of balances between density, permeability and information to materialise in the form of actual urban structures – from the urban block at the microscale to large-scale internal correlations.